\documentclass[conference,onecolumn,12pt,peerreviewca,a4paper]{IEEEtran}\usepackage[]{graphicx}\usepackage[]{color}
\makeatletter
\def\maxwidth{ %
  \ifdim\Gin@nat@width>\linewidth
    \linewidth
  \else
    \Gin@nat@width
  \fi
}
\makeatother

\definecolor{fgcolor}{rgb}{0.345, 0.345, 0.345}

\usepackage{framed}
\makeatletter
 {\par\unskip\endMakeFramed%
 \at@end@of@kframe}
\makeatother

\definecolor{shadecolor}{rgb}{.97, .97, .97}
\definecolor{messagecolor}{rgb}{0, 0, 0}
\definecolor{warningcolor}{rgb}{1, 0, 1}
\definecolor{errorcolor}{rgb}{1, 0, 0}
\newenvironment{knitrout}{}{} %

\usepackage{alltt}
\usepackage[utf8]{inputenc}

\usepackage[table,x11names,dvipsnames,table]{xcolor}
\usepackage[backend=biber]{biblatex}

\IEEEoverridecommandlockouts

\usepackage{amsmath,amssymb,amsfonts}
\usepackage{float}
\usepackage{graphicx}
\usepackage{hyperref}
\usepackage{textcomp}
\usepackage{todonotes}
\usepackage{booktabs}
\usepackage{multirow}
\usepackage{footnote}
\usepackage[doublespacing]{setspace}

\def\BibTeX{{\rm B\kern-.05em{\sc i\kern-.025em b}\kern-.08em
    T\kern-.1667em\lower.7ex\hbox{E}\kern-.125emX}}

\graphicspath{{figure/}}
\IfFileExists{upquote.sty}{\usepackage{upquote}}{}
\begin{document}

\title{Using Supervised Learning to Classify Metadata of Research Data by Discipline of Research}

\author{
	\IEEEauthorblockN{
        Tobias Weber\IEEEauthorrefmark{1}\IEEEauthorrefmark{2},
        Dieter Kranzlmüller\IEEEauthorrefmark{2},
        Michael Fromm\IEEEauthorrefmark{3},
        Nelson Tavares de Sousa\IEEEauthorrefmark{4}
    }
	\IEEEauthorblockA{
        \IEEEauthorrefmark{1}
        Corresponding Author,
        \href{mailto:weber@lrz.de}{weber@lrz.de},
        \href{https://orcid.org/0000-0003-1815-7041}
        {orcid/0000-0003-1815-7041}
    }
	\IEEEauthorblockA{
        \IEEEauthorrefmark{2}
		\textit{Leibniz Supercomputing Centre}, Garching b. München (Germany)
    }
	\IEEEauthorblockA{
        \IEEEauthorrefmark{3}
    	\textit{Database Systems Group},
    	\textit{Ludwig-Maximilians-Universität München} (Germany)\\
        \href{https://orcid.org/0000-0002-7244-4191}
        {orcid/0000-0002-7244-4191}
    }
	\IEEEauthorblockA{
        \IEEEauthorrefmark{4}
    	\textit{Software Engineering Group},
    	\textit{Kiel University} (Germany)
        \href{https://orcid.org/0000-0003-1866-7156}
        {orcid/0000-0003-1866-7156}
    }
}

\maketitle

\begin{abstract}
Automated classification of metadata of research data by their discipline(s) of research
can be used in scientometric research,
by repository service providers,
and in the context of research data aggregation services.
Openly available metadata of the DataCite index for research data
were used to compile a large training and evaluation set
comprised of 609,524 records,
which is published alongside this paper.
These data allow to reproducibly assess classification approaches,
such as tree-based models and neural networks.
According to our experiments with 20 base classes (multi-label classification),
multi-layer perceptron models perform best
with a $f_1$-macro score of
0.760
closely followed by Long Short-Term Memory models
($f_1$-macro score of 0.755).
A possible application of the trained classification models is
the quantitative analysis of trends towards interdisciplinarity of digital scholarly output or
the characterization of growth patterns of research data, stratified by discipline of research.
Both applications perform at scale with the proposed models which are available for re-use.
\end{abstract}

\IEEEpeerreviewmaketitle
\section{Introduction}\label{sec:introduction}
By research data, we understand all digital input or output of those activities of researchers
which are necessary to produce or verify knowledge in the context of the sciences and humanities \cite{8588646}.
Metadata about research data provide a common scheme
across different formats, encodings and bit-representations
and can be used as a placeholder for the research data in order to classify them.
Assigning disciplines of research to metadata of research data
is a multi-label classification problem:
research data can be mapped to multiple disciplines of research
and these disciplines are not exclusive (research data can belong to more than one discipline).
Since both the amount and the growth of research data
is too extensive to keep up with manual routines,
automatic classifiers are needed for many use cases \cite{general015}, \cite{issi004}.

Three of these use cases illustrate the usefulness of such an automated classifier
and help to specify the requirements for the data processing pipelines
and the evaluation procedures presented in this paper:

\begin{enumerate}
    \item \textbf{Scientometric Research}:
        During scientometric analyses normalization problems arise;
        the comparison of impact of publications is an example:
        to compare the values (citations, usage, etc.) across fields, these must be normalized
        to values typical for a specific field or discipline of research.
        Only automated classification procedures allow to find estimates over large data sets
        as a baseline for these normalizations.
        An automated classifier can also be used to take samples out of said large data sets,
        which are stratified by the discipline of research.
    \item  \textbf{Assistant Systems}:
        Providers of research data services can take advantage of automated classification
        in order to assist submitters to their repositories:
        Assistant systems can use the classifiers to suggest labels to the submitters
        based on the metadata submitted.
        This not only eases the work of the submitters and curators,
        but improves the overall quality of available metadata.
    \item  \textbf{Value-adding Services}:
        Research data aggregation services, such as
        DataCite\footnote{\url{https://datacite.org}} or
        BASE\footnote{\url{https://base-search.net}}
        collect metadata across different disciplines of research.
        Enriching the collected metadata by adding classification information enables
        value-adding services such as a faceted search,
        publication alarms for specific fields,
        or other discipline-specific services.
\end{enumerate}

The requirements derived by each of these three use cases differ:
using the automated classifier in sampling research data,
a wrongly assigned label has a greater impact on the application compared to a missed label.
Assistant systems on the other hand ideally identify all correct labels;
since humans can correct the suggestions in this context,
a wrongly assigned label is not as bad as a missed label.
The former use case therefore stresses the precision of the classifier,
while the latter use case stresses its recall.
In the value-adding services use case both qualities are equally important.

The general idea to realize such a classifier as proposed by this paper,
is to retrieve openly available metadata for research data,
extract title, description and keywords from the metadata,
vectorize the texts,
and find a machine learning algorithm that can predict the discipline of research.
To achieve that we used the multi-label-enabled classifiers  provided by the scikit-learn framework \cite{scikit-learn},
and neural networks as supported by tensorflow \cite{tensorflow}.
The scikit-learn framework only includes well-established models and algorithms,
which will help to improve the comparability and reproducibility of our findings.
Tensorflow offers hardware support for graphic processing units (GPUs) to speed up the training of neural networks;
the resulting neural networks can be integrated into scikit-learn's evaluation procedure
with minimal adjustments.

The main contributions of this paper are:
\begin{itemize}
    \item A methodically strict evaluation of selected classification algorithms
        for each of the use cases
    \item The publication of the data set used for this evaluation,
        which allows others to reproduce or supersede our findings
    \item The publication of the complete source code used
        to clean the data and map them to our base classification scheme
    \item Suggestions on how automated classifiers can be used for scientometric research
\end{itemize}

The remainder of this paper is structured as follows:
\autoref{sec:relatedWork} discusses the relations of our approach to a selection of published work.
In \autoref{sec:dataHandling} we summarize how the data set
was retrieved and cleaned.
The combinations of models and vectorization approaches
are presented in \autoref{sec:featuresAndModels}.
The methodological approach to evaluate these combinations is described in \autoref{sec:methods}.
\autoref{sec:results} gives an overview over the results,
i.e.\ the best approaches for the use cases.
These findings are discussed in \autoref{sec:discussion},
which includes a comment on threats to validity.
The last section concludes and gives an outlook including suggestions for exploiting
our findings in order to answer scientometric questions.
\section{Related Work} \label{sec:relatedWork}

An introduction to multi-label classification can be found in \cite{Sorower} and \cite{multilabelML006}, an overview and discussion of metrics is presented in \cite{fscore}.
This section lists the most promising approaches to achieve
an automated classification of research data by discipline of research and their common shortcomings,
then discusses the available data to evaluate the approaches and concludes
with the presentation of work discussing technical specifics of multi-label classification of textual data
that were of interest for our approach.

\subsection{Automatic Classification by Discipline of Research}

The approach presented in \cite{ddcsvm} uses a Support Vector Machine (SVM) model
to classify according to the Dewey Decimal Classification (DDC, in three hierarchy levels).
The problem is described as a multi-label classification task
with (partially) hierarchical predictions.
The partial nature is due to the sparseness of the used training set,
which was compiled out of the data available via the BASE service (Bielefeld Academic Search Engine)\footnote{\url{https://base-search.net}} at that time:
On the first level of the DDC hierarchy 5,868 English and 7,473 German metadata records
were available,
on the second and third level 20,813 English and 37,769 German metadata records
were available.
The English classifier had a $f_1$-score of 0.81 (classification over base classes, i.e.\ 10 labels);
for the deeper levels only partial data are available.
It is not specified, whether the score is averaged over the data set (micro)
or the scores of each label (macro).
In comparison to \cite{ddcsvm} our training set is approximately 30 times larger.

The authors of \cite{ddcClass} also discuss the possibilities to apply machine learning algorithms
on bibliographic data labeled with DDC numbers.
They evaluate their approach with a data set comprising of classical publications
limited to the DDC classes 500 and 600 (science and technology) with 88,400 mostly single-labeled records.
The authors suggest to flatten the hierarchy to reduce the initially high number of labels (18.462).
Although the proposed classification approach achieves an accuracy score of nearly 90\%,
it is no viable option for our use cases,
since it is based on a multi-class approach
(classification over multiple labels, which are taken as exclusive)
and interactions of humans.
As already stated, the approach is limited to a relatively narrow selection of disciplines of research.

Another approach to use SVM models to predict DDC research disciplines is presented in \cite{rw001}.
The authors characterize the problem as multi-class, but their classifier honors the hierarchy of DDC.
The used data set includes 143,838 records from the Swedish National Union Catalogue (joint catalogue of the Swedish academic and research libraries).
They report a peak accuracy of 0.818.

In general, classifiers targeting DDC (\cite{ddcsvm}, \cite{ddcClass}, \cite{rw001}),
face the problem that predicting the first DDC level is typically not very useful
(only 10 classes, one of which is "Science"),
whereas classifiers targeting DDC's second level need to provide reliable results for 100 labels;
for the latter task to succeed the reported data sets are too small
and results are as a consequence partial at best.
Our analysis of the DataCite index furthermore indicates that DDC
is not necessarily the most used classification scheme for research data,
despite its popularity among information specialists and librarians
(see \autoref{tab:schemes}).
These problems could be circumvented by using a classification scheme that is expressive enough in the first level,
as proposed by us.
As a conclusion of the review of the literature we furthermore decided
to not include hierarchy predictions to our problem.
We understand the classification problem at hand as twofold:
Classifying base classes on the one hand and determine the depth in the hierarchy on the other.
The latter could itself be understood (recursively) as a multi-label classification problem.
While we hope to contribute to the former, we do not claim to solve the latter.

All approaches that we found in the literature have at least one of the following shortcomings:
\begin{itemize}
    \item The classification task is characterized as multi-class, not multi-label.
    \item The reported classification performance is not comparable to other approaches,
        since important values are missing or reported values are too unspecific.
    \item The domain of classification only includes classical publications as opposed to the more general class of research data.
    \item The evaluation of the approaches is limited to a subset of the possible base classes or labels.
    \item The classification routine includes human interaction.
\end{itemize}

Our approach shares none of the named shortcomings.
The literature furthermore concentrates on linear machine learning models (most prominently SVMs),
which is why we excluded them from the evaluation and
concentrated on tree-based models and neural networks instead.

\subsection{Data Publications to Evaluate Classification Approaches}

A general problem we found in the course of reviewing the available literature
is the in-comparability of the reported results;
incomplete or in-commensurable performance metrics are not the only issue:
The values could have been re-calculated if the data of the publication were available.
With one exception (\cite{unsup}),
all publications we found do not include enough information
to retrieve the data used to evaluate the presented approach.
Additionally, different data sets might lead to different results;
there is no single, canonical data set
which is used to evaluate the different approaches.

In \cite{corpus} an approach is presented to compile an annotated corpus of metadata
based on the OAI-PMH standard, the Dublin Core metadata scheme and the DDC classification scheme.
The authors created a manual mapping to determine the DDC label.
The resulting data set includes 52,905 English records annotated with one of the 10 top-level DDC classes.
We improved their approach by using the DataCite metadata scheme \cite{datacite43}
which supports qualified links to classification schemes.
This allowed us to compile a larger data set with a finer set of base classes
and the possibility to integrate different classification schemes into our approach.
We found the resulting data set in a similar imbalance as
the data set presented by \cite{corpus}
(cp. \autoref{sec:dataHandling}).

We hope to contribute not only with our classification approach but also by providing a large data set
that can be used to evaluate future approaches and reproduce the findings of already proposed approaches.

\subsection{Multi-label Classification of Textual Data}

The authors of \cite{multilabelML001} propose a multi-label classification approach of social media texts based on
a combination of a graph-based method with a semi-supervised approach.
The problem of social media multi-label classification is similar to our problem,
since both handle classification of short and heterogeneous texts with diverse creators.
However, approaches to select algorithms to classify, clean and prepare data differ.
The reason lies in the different domains of the classification problems:
While social media data are linked via hashtags or mentions and texts tend to be written in an informal tone
or even in a particular slang,
both description and linking (citations) in our use cases are more formal.
This suggests that a less sophisticated approach than presented in \cite{multilabelML001}
might still provide satisfactory results.
A similar line of reasoning applies to \cite{multilabelML015} which is based on Latent Dirichlet Allocation.

Stratified sampling is a necessary step in typical multi-label classification pipelines.
In \cite{multilabelML014} an algorithm to realize a
"relaxed interpretation of stratified sampling for multi-label data" is proposed.
The base idea is to distribute the data items over n subsets,
starting with all data items labeled with the least common label (greedy approach).
Since our data includes a substantial part with only one label, we followed another
approach which is easier to implement (see \autoref{sec:dataHandling} for details).

\cite{nothman2018stop} voice concerns with regard to stop word lists in vectorizing text data
(controversial words, incompatible tokenization rules, incompleteness).
Additionally, the contextuality of stop word lists are a problem
which is not elaborated in depth by these authors:
If the context of a set of documents is given,
certain words are likely to lose discriminatory potential,
although they would not qualify as stop words in a more general context.
We decided to extend an existing stop word list (see \autoref{sec:dataHandling} for details),
to take advantage of the context of data.

\section{Data Handling}\label{sec:dataHandling}
\subsection{Data Retrieval}\label{subsec:retrieve}

The training and evaluation data have been retrieved from
the DataCite index of metadata of research data\footnote{\url{https://datacite.org}}
via OAI-PMH.\footnote{\url{http://www.openarchives.org/OAI/openarchivesprotocol.html}}
DataCite is a service provider that aggregated research data over more than 1,100 publishers and 750 institutions in 2017 \cite{issi005}.
In September 2019 more than 18.75 million metadata records for research data were available on DataCite's index.
DataCite is also the name of a metadata schema \cite{datacite43}.
All metadata available via OAI-PMH comply to one of the versions of this scheme.

The retrieved data include metadata from June 2011 to May 2019.
We used a customized GeRDI-Harvester\footnote{Generic Research Data Infrastructure, \url{https://www.gerdi-project.de}}
to retrieve the metadata in DataCite format and to filter out any non-qualified records;
a qualified record is understood as a metadata record with
at least one subject field that is qualified either with an URI to a scheme (schemeURI attribute)
or a name for a scheme (subjectScheme attribute).

In sum, 2,476,959 metadata records are the input of the cleaning step (see following paragraph).
The data retrieval took place in May 2019.
The total number of items in the index at this time was approximately 16 million records.\footnote{
The exact number is not available due to the length of the time span the harvesting process took,
in which ingests to and deprovisionings from the DataCite index took place.}

Ideas to selectively include additional sources for underrepresented disciplines of research
were discarded, since these would bias the data to metadata labeled with only one label:
all data sources specific to a certain discipline, did not support multi-labeling.

\begin{knitrout}
\definecolor{shadecolor}{rgb}{0.969, 0.969, 0.969}\color{fgcolor}\begin{table}[!h]

\caption{\label{tab:schemes}Supported Classification Schemes}
\centering
\resizebox{\linewidth}{!}{
\begin{tabular}{lr}
\toprule
Name & Records (after cleaning)\\
\midrule
\rowcolor{gray!6}  Australian and New Zealand Standard Research Classification (ANZSRC) & 374,472\\
Dewey Decimal Classification & 212,352\\
\rowcolor{gray!6}  Digital Commons Three-Tiered Taxonomy of Academic Disciplines & 11,674\\
Basisklassifikation & 7,032\\
\rowcolor{gray!6}  Narcis Classification Scheme & 4,015\\
\bottomrule
\multicolumn{2}{l}{\textit{Note: } n = 609,524; a record can be qualified by more than one scheme}\\
\end{tabular}}
\end{table}

\end{knitrout}
\subsection{Data Cleaning}\label{subsec:clean}

\subsubsection{A Common Classification Scheme}

Six classification schemes for disciplines of research are frequently used throughout
the retrieved metadata, our method supports five of them
(cp.\ \autoref{tab:schemes}).
The scheme missing from the table is linsearch, which is a classification scheme that is derived from automatic classification (see \cite{ddcsvm},\cite{linsearch}).
We decided to exclude all instances of metadata which we could identify as automatically labeled
to avoid amplifier effects;
data sets which were classified by machine learning algorithms
are necessarily biased towards the algorithm used
and what a machine can classify in general.

In the course of cleaning the data ,
a common classification scheme was defined
(see \autoref{tab:labels}),
which closely resembles the most common scheme,
the Australian and New Zealand Standard Research Classification (ANZSRC).
Two pairs of classes were merged in order to map the other classification scheme
to the common classification scheme:
\begin{itemize}
    \item "Earth Sciences" and "Environmental Sciences" are divisions 04 and 05 resp.\
        of the ANZSRC classification scheme
        and became "Earth and Environmental Sciences"
        in the common classification scheme.
    \item "Engineering" and "Technology" are divisions 09 and 10 resp.\
        of the ANZSRC classification scheme
        and became "Engineering and Technology"
        in the common classification scheme.
\end{itemize}

These merges enabled a mapping from the other classification schemes to ANZSRC
without arbitrary splits or losing records due to miss-matches of the schemes.
The resulting classification scheme has been flattened (projection to the base classes)
and has therefore no hierarchy.
The exact mappings of the schemes to the common classification scheme is available for analysis
and improvement (cp. \autoref{subsec:reproducibility}).

\subsubsection{Cleaning Procedure}

\begin{knitrout}
\definecolor{shadecolor}{rgb}{0.969, 0.969, 0.969}\color{fgcolor}\begin{table}[!h]

\caption{\label{tab:labels}Base Classes and Occurrences in the Data Set}
\centering
\resizebox{\linewidth}{!}{
\begin{tabular}{lrrrrrrrrr}
\toprule
Class & 1 label & 2 labels & 3+ labels & best & total & \% & $\varnothing$\#labels & $\varnothing$wc & wc (med.)\\
\midrule
\rowcolor{gray!6}  Mathematical Sciences & 9,144 & 13,635 & 23,719 & 45,925 & 46,498 & 7.63 & 2.55 & 111 & 59\\
Physical Sciences & 130,593 & 8,556 & 13,420 & 130,593 & 152,569 & 25.03 & 1.27 & 50 & 22\\
\rowcolor{gray!6}  Chemical Sciences & 16,349 & 27,090 & 37,958 & 57,086 & 81,397 & 13.35 & 2.44 & 141 & 105\\
Earth and Environmental Sciences & 13,369 & 24,754 & 35,355 & 57,075 & 73,478 & 12.05 & 2.48 & 144 & 97\\
\rowcolor{gray!6}  Biological Sciences & 67,884 & 88,169 & 71,194 & 86,325 & 227,247 & 37.28 & 2.11 & 124 & 66\\
Agricultural and Veterinary Sciences & 1,876 & 892 & 431 & 3,164 & 3,199 & 0.52 & 1.63 & 141 & 76\\
\rowcolor{gray!6}  Information and Computing Sciences & 27,723 & 15,159 & 27,091 & 51,472 & 69,973 & 11.48 & 2.17 & 115 & 74\\
Engineering and Technology & 25,104 & 6,449 & 2,202 & 29,850 & 33,755 & 5.54 & 1.35 & 165 & 146\\
\rowcolor{gray!6}  Medical and Health Sciences & 68,121 & 46,737 & 42,678 & 86,325 & 157,536 & 25.85 & 1.93 & 134 & 77\\
Built Environment and Design & 1,800 & 1,108 & 360 & 3,183 & 3,268 & 0.54 & 1.61 & 147 & 86\\
\rowcolor{gray!6}  Education & 2,499 & 1,341 & 1,262 & 4,913 & 5,102 & 0.84 & 1.91 & 124 & 99\\
Economics & 5,211 & 1,238 & 1,119 & 6,644 & 7,568 & 1.24 & 1.62 & 151 & 133\\
\rowcolor{gray!6}  Commerce, Management, Tourism and Services & 5,128 & 1,123 & 498 & 6,217 & 6,749 & 1.11 & 1.36 & 132 & 116\\
Studies in Human Society & 6,726 & 4,203 & 1,294 & 9,303 & 12,223 & 2.01 & 1.65 & 137 & 129\\
\rowcolor{gray!6}  Psychology and Cognitive Sciences & 11,458 & 4,744 & 1,812 & 15,400 & 18,014 & 2.96 & 1.52 & 138 & 138\\
Law and Legal Studies & 1,048 & 185 & 146 & 1,338 & 1,379 & 0.23 & 1.42 & 174 & 155\\
\rowcolor{gray!6}  Studies in Creative Arts and Writing & 1,118 & 290 & 326 & 1,519 & 1,734 & 0.28 & 1.58 & 142 & 106\\
Language, Communication and Culture & 4,482 & 979 & 613 & 5,442 & 6,074 & 1.00 & 1.41 & 117 & 88\\
\rowcolor{gray!6}  History and Archaeology & 5,703 & 645 & 284 & 6,166 & 6,632 & 1.09 & 1.19 & 55 & 32\\
Philosophy and Religious Studies & 473 & 731 & 394 & 1,584 & 1,598 & 0.26 & 2.02 & 124 & 111\\
\rowcolor{gray!6}  total & 405,809 & 124,014 & 79,701 & - & 609,524 & 100.00 & 1.50 & 113 & 58\\
\bottomrule
\multicolumn{10}{l}{\textit{Note: } 2,3,4+ labels do not sum to their total; total/total is the sum of these totals.}\\
\end{tabular}}
\end{table}

\end{knitrout}

\begin{figure}
    \centering
    \includegraphics[width = \textwidth,clip]{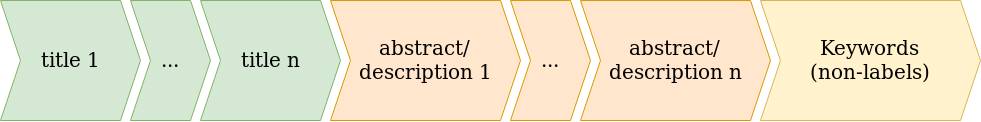}
    \caption{Concatenation of the payload parts}
    \label{fig:concat}
\end{figure}

The data were cleaned along this procedure:
\begin{enumerate}
    \item Mapping from the supported schemes (\autoref{tab:schemes}) to one or more disciplines of research
        according to the classification scheme of this paper (\autoref{tab:labels});
        after this step 1,233,427 records remained,
        1,243,532
        records were filtered out as "not annotatable",
        i.e.\ there was no mapping to a discipline of research available.
        Typical reasons for a missing mapping include
        unclear identification of the source scheme and
        different domain of the scheme (meaning it does not classify discipline(s) of research).
    \item Filter out duplicates;
        after this step 716,180 records remained,
        517,247 records
        were filtered out as duplicates.
    \item Creation and validation of the payload;
        the payload consists of one ore more titles,
        zero or more abstracts/descriptions and a subset of the subjects of the research data.
        This subset consists of all subject tags which have \emph{not} been used for the mapping to
        the target classification scheme and can be empty.
        The order of concatenation is depicted in \autoref{fig:concat}.
        Only those parts are concatenated with each other which consists mostly of English words.\footnote{\
            We used a python port of the langdetect library \cite{langdetect} to determine the language of the fields:
            \url{https://pypi.org/project/langdetect}
        }
        If the resulting concatenated string is less than 10 words (separated by white space),
        it was discarded.
        After this step 609,524 records remained,
        106,656 were filtered out as not fitting for the purpose.
\end{enumerate}

\autoref{tab:labels} provides some statistics for the resulting data set:
\begin{itemize}
    \item \textbf{1/2/3+ label(s)}:
        all metadata records with exactly one, two, and three or more labels
    \item \textbf{best}:
        all metadata records for which this discipline of research
        is the best label for stratification (see below)
    \item \textbf{total}:
        all metadata records labeled with this discipline of research
        (the sum of all these values is bigger than the number of clean records, since a record can have more than one label)
    \item \textbf{\%}:
        percentage of metadata records labeled as this discipline of research,
        rounded to two decimal positions.
    \item \textbf{$\varnothing$\#labels}:
        arithmetical mean of the number of labels per record with that label (rounded to two decimal digits) .
    \item \textbf{$\varnothing$wc}:
        arithmetical mean of the number of words per record with that label (rounded to an integer).
    \item \textbf{wc (med.)}:
        median of the number of words per record with that label.
\end{itemize}

The distribution of the different disciplines of research is imbalanced,
which is in accordance with the findings in the literature \cite{ddcsvm}, \cite{rw001},
\cite{issi001}, \cite{issi003}.
The imbalance of labels necessitates additional thought on
the selection of evaluation metrics in \autoref{sec:methods}
and on the configuration of the different models.

Payloads labeled as "Physical Sciences" and "History and Archaeology"
are noteworthy outliers,
since they have fewer words compared to records from other disciplines of research.
The average number of labels is highest in "Mathematical Sciences",
since statistics is part of this category,
which is a field that is intertwined with disciplines using quantitative methods.
With one exception all disciplines have a distribution of word counts that is skewed to shorter payloads
(median < mean), meaning that there are length-wise outliers.
Psychology shows the same mean as median.

The label cardinality (average number of labels per record) is 1.5,
the label density (average proportion of labels per record) is 0.08.\footnote{
    see \cite{multilabelML006} for a formal definition}
The data set includes 961 different labelsets
this number is relatively small compared to the $20^{20}$ theoretically possible labelsets.
179 labelsets occur only once which makes a stratified split along the 961 labelsets impossible.
To enable stratified splitting we followed a "best-label" approach:
\begin{enumerate}
    \item All records with only one label are assigned that label.
    \item Iterating over the remaining records, the "best" label out of the labelset is selected,
        which is the label which is selected the least often at the current state of the  loop.
\end{enumerate}
These best labels are used in stratified sampling and feature selection (see following paragraph),
but \emph{not} as labels for the training itself.

\section{Vectorization and Model Selection}\label{sec:featuresAndModels}

The general idea is to evaluate three different classes of combinations for the use cases presented in \autoref{sec:introduction}:
\begin{enumerate}
    \item Classic machine learning models combined with bag of words vectorization;
       Approaches which have already been evaluated (linear models)
        or which do not support multi-label classification natively have been excluded.
    \item Classic deep learning models (multi-layer perceptrons),
        combined with bag of words vectorization
    \item A model used in contemporary Natural Language Processing,
        the Long Short Term Memory (LSTM) model,
        combined with word embedding vectorization.
\end{enumerate}

This section first describes the two vectorization approaches;
the second part presents the machine learning models, which are the evaluation candidates of \autoref{sec:results}.

\subsection{Vectorization}\label{subsec:vectorize}
Before the vectorization, the data are split into a training (548,571 records)
and an evaluation set (60,953 records) with a ratio of 9:1.
The split is stratified, i.e.\ the distribution of the best labels in test and training set
is approximately identical.
For the deep learning models the training set is again split by the same approach (493,713 training and 54,858 validation records).
Both vectorization methods operate on the same splits to gain comparable results.
The split and the vectorization is executed three times,
for a small (s), medium (m) and large (l) vectorized representation of the payloads;
the definition of the sizes will be given in the following paragraphs.

\subsubsection{Bag of Words (BoW)}
One way to vectorize the corpus of all documents,
i.e.\ all payloads of the metadata records,
is the "Bag of n-grams"-approach,
i.e.\ each document is treated as a row in a matrix
in which the columns are the terms (1-grams and 2-grams).
Some terms were filtered out by a stop word list.
A stop word list is designed to filter out non-informative parts of the documents.
We chose to create our own stop word list (with 240 entries,
see \autoref{sec:relatedWork} for a rationale).
The list of stop words includes:
\begin{itemize}
    \item words that are generally considered unspecific
        (e.g.\ "the", "a", "is")\footnote{
            These stop words are a subset of the English stop words list
            of the nltk software package \cite{nltk}}
    \item numbers and numerals $\leq$ 10
    \item words that are unspecific in the context of data
        (e.g.\ "kb", "file", "metadata", "data")
    \item words that are unspecific in the academic context
        (e.g.\ "research", "publication", "finding")
\end{itemize}

For each term $t$ in a document,
that is for each cell of the matrix of documents and terms,
the term frequency-inverted document frequency (tf-idf)
was calculated using the default settings of scikit-learn \cite{scikit-learn} (except for the stop word list).
The vectorization resulted in 4,096,093 possible features.

The best features were selected in three modes:
\begin{itemize}
    \item \textbf{s}: 1000 features per label, 20,000 features in total
    \item \textbf{m}: 2500 features per label, 50,000 features in total
    \item \textbf{l}: 5000 features per label, 100,000 features in total
\end{itemize}

The selection is based on an ANalysis Of VAriance (ANOVA) of the features \cite{fisher}.
This allows to identify the features which are best suited to discriminate between the classes.
This is the second and last time that the best labels were used.

\subsubsection{Word embeddings}
\label{sec:embeddings}
Another approach for vectorization is using word representations,
like word2vec \cite{word2vec}.
Such an approach can utilize either continuous bag-of-words (CBOW) \cite{cbow} or continuous skip-gram \cite{word2vec}.
The CBOW model predicts the current word from a window spanning over context words.
The skip-gram model uses the current word to predict the context surrounding the word.
In our work we used pre-trained word2vec embeddings\footnote{https://code.google.com/archive/p/word2vec/} which were
trained on the Google News data set (about 100 billion words).
The embeddings were trained with the CBOW approach and consist of 300-dimensional word vectors representing three million words.
Compared to BoW, word embeddings provide a low dimensional feature space
and encode semantic relationship among words.

To apply the vectorization method, each document needs to be tokenized:
\begin{itemize}
    \item \textbf{s}: up to 500 words of the document are tokenized
    \item \textbf{m}: up to 1,000 words of the document are tokenized
    \item \textbf{l}: up to 2,000 words of the document are tokenized
\end{itemize}

\subsection{Multi-label Models}

\subsubsection{Classic Machine Learning Models}
\begin{itemize}
    \item \emph{DecisionTreeClassifier} \cite{dct}:
        This classifier uses a decision tree to find the best-suited classes for each record.
        The nodes of the tree are used to split the records into two sets based on a feature,
        eventually resulting in a leaf which ideally represents a certain class or a certain labelset.
        The training consists in finding the best features to split the data set.
        Multi-label classification of unseen data works by following the decision path
        until a leaf is reached.
        All labels which are present in the majority of the data items in the leaf are returned
        as the classification result.
    \item \emph{RandomForestClassifier} \cite{rf}:
        This classifier is based on the DecisionTreeClassifier,
        by building an ensemble of multiple decision trees.
        The general idea is that a bias that trees typically have by overfitting a training set,
        is remedied by building multiple decision trees based on different
        random subsets of the features.
        In ideal cases the bias in different directions corresponds to different aspects of the data.
        The training consists in fitting n trees by using a random subset of features
        and a random subset of the training set.
        The classification is then achieved by a voting procedure among the trees in the ensemble.
    \item \emph{ExtraTreesClassifier} \cite{et}:
        This classifier is similar to the RandomForestClassifier: both are ensembles of trees,
        but this classifier is based on the ExtraTreeClassifier (note the missing "s") which introduces more randomness
        by selecting the feature to split by totally at random.
        (DecisionTreeClassifiers in random forests
        by contrast select the best feature out of a subset sampled at random).
\end{itemize}

\subsubsection{Multi-Layer Perceptron}
A Multi-Layer Perceptron (MLP) \cite{mlp} is a neural network
and consists of one input layer, one output layer and n intermediate layers of perceptrons.
The input layer corresponds to the vectorized data
(e.g. a vector of 20,000 values in the s-sized BoW approach)
the output layer has the shape of the labels (i.e.\ a vector with 20 elements).
Backpropagation is used to train the the model for the given data;
we used the Adam optimizer for this task \cite{adam}.
By using a sigmoid function as the activation function of the output layer,
the MLP can predict the probability for multiple labels (multi-label).

\subsubsection{Recurrent Network}
Recurrent networks form a class of neural networks which are used to process sequential data of different length. This design furthermore
allows it to make use of temporal dynamic behavior,
e.g.\ recurrences of terms in a text.
The recurrent network architecture we use in our work is an Bidirection Long Short-Term Memory (BiLSTM) \cite{lstm} model.
We use word2vec \cite{word2vec} embeddings as described in Section \ref{sec:embeddings} as input to the model.\footnote{Any other kind of word embeddings can be used too}
The embeddings are frozen and not further trained in the classification process.
On top of the BiLSTM layer we use a dense layer with a sigmoid activation function
to classify multi-labels.
The BiLSTM layer and the dense layer get trained by an Adam optimizer \cite{adam}.

\subsubsection{Weights and Hyper-parameter Tuning}
An important parameter in the training of multi-label classification problems based on imbalanced
training sets is the weight given to each label.
All classes of algorithms we used allow to put more weight to underrepresented labels.
We calculated the weights based on the label frequencies found in the training set:
    \[\text{weight(label)} = \frac{1}{\frac{\text{frequencies(label})}{\text{max(frequencies)}}}\]

For each model we executed semi-automated parameter tuning,
following this procedure:
\begin{enumerate}
    \item List selected hyper-parameters in the order of expected impact to the evaluation metrics.
    \item For each hyper-parameter or combination thereof, execute a grid search to find the best candidate(s).
    \item Fix the selected parameters and repeat step 2 with the next parameter or parameter combination.
\end{enumerate}
The space of possible solutions is too big to be exhaustively searched with reasonable efficiency,
so it might be, that a different combination of hyper-parameters improve the scores reported by us.

\section{Evaluation procedure}\label{sec:methods}

For each use case (cp.\ \autoref{sec:introduction}) an evaluation metric has been identified
that takes the imbalance of the label distribution into account, as stated in \autoref{sec:dataHandling}.
The use cases differ in the weight they put on recall and precision.
Recall for label $l$ is the ratio between true positives and positives for label $l$
(sum of true positives and false negatives for $l$).
Precision for label $l$ is the ratio between true positives and predicted positives
(sum of true positives and false positives for $l$).\footnote{formal definitions can be found in \cite{Sorower}}

These values alone are easy to game, which is why they should be combined:
The $f_{\beta}$-score puts them in a relation to each other
that allows to modify the weight we put to precision and recall respectively:
\[f_{\beta} = (1 + \beta^2) \cdot \frac{\text{precision} \cdot \text{recall}}{\beta^2 \cdot \text{precision} + \text{recall}}\]

The value of $\beta$ controls the weight put to recall and precision:
\begin{itemize}
    \item If $\beta < 1$, precision is highlighted; a value of 0.5 has been chosen for the "scientometric research" use case.
    \item If $\beta > 1$, recall is highlighted; a value of 2 has been chosen for the "assistant system" use case.
    \item If $\beta = 1$, precision and recall are treated equally; this is chosen for the "value-adding services" use case.
\end{itemize}

Precision and recall is calculated for each label,
and the arithmetical mean over all labels is taken as the input
for the calculation of the $f_{\beta}$-score.
This \emph{macro}-average approach takes the imbalance of the base classes
(and therefore of the labels) into account.
It can be interpreted as the chance of a correct classification when a stratified sample is drawn.
This is the basis for the evaluation of the approaches for the presented use cases.

The \emph{micro}-average approach averages the values over all data sets, without the intermediate
aggregation over the labels.
It can be interpreted as the chance of a correct classification
when a completely random sample is drawn.
In imbalanced scenarios, micro-scores tend to be skewed to the predominant labels;
since these often perform better (more training data), micro-scores are often too optimistic,
when the performance of the model with regard to all labels is the target.
Although micro-averages are not used in this paper to evaluate the models,
we nevertheless report them for the sake of comparability.

We refrain from reporting accuracy
(ratio of the sum of true positives and true negatives for label $l$
to the size of the evaluation set),
since it is biased towards negative classifications
which in our case are much more frequent than positive classifications.

The final evaluation is based on the $f_{\beta}$-macro-scores
calculated on all three evaluation sets.
This way each model is tested against the same unseen set of data.
(see \autoref{sec:dataHandling}).

\section{Results}\label{sec:results}

\subsection{Model Performance}

\begin{knitrout}
\definecolor{shadecolor}{rgb}{0.969, 0.969, 0.969}\color{fgcolor}\begin{table}[!h]

\caption{\label{tab:models}Best Model Performances (aggregated)}
\centering
\resizebox{\linewidth}{!}{
\begin{tabular}{llrrrrrr}
\toprule
Model & Size & $f_{0.5}$ (macro) & $f_{0.5}$ (micro) & $f_1$ (macro) & $f_1$ (micro) & $f_2$ (macro) & $f_2$ (micro)\\
\midrule
\rowcolor{gray!6}  LSTMClassifier & s & 0.788 & 0.846 & 0.737 & 0.826 & 0.695 & 0.808\\
LSTMClassifier & m & 0.791 & 0.854 & 0.755 & 0.839 & 0.723 & 0.826\\
\rowcolor{gray!6}  LSTMClassifier & l & 0.772 & 0.832 & 0.717 & 0.809 & 0.671 & 0.787\\
MLPClassifier & s & 0.784 & 0.841 & 0.747 & 0.833 & 0.719 & 0.828\\
\rowcolor{gray!6}  MLPClassifier & m & 0.795 & 0.854 & 0.754 & 0.841 & 0.720 & 0.829\\
MLPClassifier & l & 0.809 & 0.858 & 0.760 & 0.847 & 0.729 & 0.842\\
\rowcolor{gray!6}  RandomForestClassifier & s & 0.632 & 0.779 & 0.463 & 0.667 & 0.378 & 0.584\\
RandomForestClassifier & m & 0.551 & 0.727 & 0.405 & 0.599 & 0.333 & 0.510\\
\rowcolor{gray!6}  RandomForestClassifier & l & 0.532 & 0.716 & 0.382 & 0.577 & 0.311 & 0.483\\
DecisionTreeClassifier & s & 0.296 & 0.507 & 0.321 & 0.524 & 0.369 & 0.543\\
\rowcolor{gray!6}  DecisionTreeClassifier & m & 0.281 & 0.489 & 0.303 & 0.506 & 0.351 & 0.527\\
DecisionTreeClassifier & l & 0.276 & 0.483 & 0.297 & 0.499 & 0.344 & 0.516\\
\rowcolor{gray!6}  ExtraTreesClassifier & s & 0.504 & 0.699 & 0.359 & 0.545 & 0.291 & 0.447\\
ExtraTreesClassifier & m & 0.468 & 0.681 & 0.327 & 0.513 & 0.263 & 0.411\\
\rowcolor{gray!6}  ExtraTreesClassifier & l & 0.451 & 0.667 & 0.310 & 0.491 & 0.246 & 0.388\\
\bottomrule
\end{tabular}}
\end{table}

\end{knitrout}
For all three use cases the MLPClassifier trained on the l-sized data
was the best performing model according to our tests
(cp. \autoref{tab:models}).
LSTM models trained on the m-sized data were almost as well-performing,
but took essentially longer to train than the MLP models.
The slight lead of the MLPClassifier might be explained by the amount of short payloads
in the data set:
the median word count (58 words) is left of the mean (112.78 words),
and 26 words is the value of the 25th percentile
(25 \% of the records were at most 26 words long, with 10 words being the minimum).
Many of the records' payloads might be too short for the LSTM model
to play out its ability to detect semantic relationships beyond the
statistical approach used by the MLP based on BoW.

The results of tree models  are out-of-reach of the results achieved by the deep learning models.
Trees and Ensembles perform best on s-sized data and with the exception of simple DecisionTreeClassifiers,
all models performed better in terms
of precision than in recall ($f_{0.5}$ scores are greater than $f_2$ scores).

\subsection{Performance by Discipline}

\begin{knitrout}
\definecolor{shadecolor}{rgb}{0.969, 0.969, 0.969}\color{fgcolor}\begin{table}[!h]

\caption{\label{tab:disc_scores}F-scores for each Discipline of Research (best MLP-l/LSTM-m-model)}
\centering
\resizebox{\linewidth}{!}{
\begin{tabular}{lllllll}
\toprule
Model & $f_{0.5}$-MLP & $f_1$-MLP & $f_2$-MLP & $f_{0.5}$-LSTM & $f_1$-LSTM & $f_2$-LSTM\\
\midrule
\rowcolor{gray!6}  Mathematical Sciences & 0.79 & 0.76 & 0.74 & 0.77 & 0.74 & 0.71\\
Physical Sciences & 0.96 & 0.95 & 0.94 & 0.95 & 0.94 & 0.92\\
\rowcolor{gray!6}  Chemical Sciences & 0.83 & 0.82 & 0.81 & 0.81 & 0.81 & 0.81\\
Earth and Environmental Sciences & 0.79 & 0.78 & 0.76 & 0.81 & 0.78 & 0.74\\
\rowcolor{gray!6}  Biological Sciences & 0.88 & 0.89 & 0.89 & 0.88 & 0.88 & 0.89\\
Agricultural and Veterinary Sciences & 0.70 & 0.58 & 0.50 & 0.68 & 0.60 & 0.54\\
\rowcolor{gray!6}  Information and Computing Sciences & 0.82 & 0.81 & 0.79 & 0.81 & 0.79 & 0.77\\
Engineering and Technology & 0.80 & 0.77 & 0.76 & 0.78 & 0.76 & 0.74\\
\rowcolor{gray!6}  Medical and Health Sciences & 0.83 & 0.83 & 0.81 & 0.85 & 0.83 & 0.82\\
Built Environment and Design & 0.75 & 0.64 & 0.58 & 0.71 & 0.63 & 0.57\\
\rowcolor{gray!6}  Education & 0.78 & 0.73 & 0.69 & 0.74 & 0.70 & 0.67\\
Economics & 0.74 & 0.69 & 0.64 & 0.75 & 0.70 & 0.65\\
\rowcolor{gray!6}  Commerce, Management, Tourism and Services & 0.71 & 0.66 & 0.61 & 0.72 & 0.67 & 0.63\\
Studies in Human Society & 0.77 & 0.73 & 0.70 & 0.76 & 0.74 & 0.71\\
\rowcolor{gray!6}  Psychology and Cognitive Sciences & 0.84 & 0.82 & 0.81 & 0.84 & 0.82 & 0.80\\
Law and Legal Studies & 0.81 & 0.72 & 0.66 & 0.83 & 0.77 & 0.73\\
\rowcolor{gray!6}  Studies in Creative Arts and Writing & 0.84 & 0.75 & 0.69 & 0.71 & 0.65 & 0.60\\
Language, Communication and Culture & 0.80 & 0.73 & 0.72 & 0.78 & 0.75 & 0.72\\
\rowcolor{gray!6}  History and Archaeology & 0.93 & 0.88 & 0.87 & 0.90 & 0.87 & 0.84\\
Philosophy and Religious Studies & 0.82 & 0.68 & 0.60 & 0.75 & 0.68 & 0.61\\
\bottomrule
\multicolumn{7}{l}{\textit{Note: } $f_1$-MLP and $f_2$-MLP are from the same model, as are all LSTM scores}\\
\end{tabular}}
\end{table}

\end{knitrout}
\autoref{tab:disc_scores} lists the un-aggregated scores for each discipline of research.
The $f_{0.5}$ score correlates positively with the amount of records (total in \autoref{tab:labels}):
0.518 (Pearson correlation).
None of the disciplines of research with more than 10,000 labeled payloads scored
a smaller $f_{0.5}$ value than 0.79,
while on the other side of the scale (less than 5,000 labeled payloads) no comparable tendency could be detected.

The LSTMClassifier scores slightly better or comparable to the MLPClassifier in disciplines of research often clustered as "life sciences",
while it performs clearly worse in comparison to the MLPClassifier in the humanities.

\subsection{Use Cases}

\subsubsection{Scientometric Research}
The results for this use case show the best scores compared to the other use cases.
The MLPClassifier is the best-performing model.  On the basis of the scores, confidence considerations can be implemented,
which allow to quantify the expected error in the classification task.
The MLPClassifier can be expected to perform best,
if the task at hand is specific to a certain discipline (e.g.\ singling-out Physics).

\subsubsection{Assistant Systems}
With the exception of "Biological Sciences", all un-aggregated values for the $f_2$-scores
are smaller than the $f_{0.5}$-scores.
Analyzing the un-aggregated scores by discipline of research, some drop more than others.
One of the common features of those disciplines is their relatively low number of total payloads.
The Pearson correlation between the total number of payloads per label and the corresponding $f_{\beta}$-scores,
gets stronger, the more weight is laid on recall:
\begin{itemize}
    \item $f_{0.5}$ 0.518
    \item $f_1$ 0.695
    \item $f_2$ 0.698
\end{itemize}
The effect of the imbalance of the training and evaluation set is therefore smaller on precision as it is on recall.
Assistant systems based on the proposed models are possible,
although their acceptance by users seems doubtful,
if they fail to suggest obvious labels.

\subsubsection{Value-adding Services}
Unsurprisingly, the $f_1$-scores of the best models lie between their neighboring extremes,
but slightly closer to the $f_2$-scores than to the $f_{0.5}$-scores.
This is due to the fact that the $f_1$-score is the harmonic mean between recall and precision,
which tends to stress the lower values.
Analogous to the previous use case, value-adding services based on the proposed model
should be tested by interaction studies whether users accept their performance.

\section{Discussion}\label{sec:discussion}

\subsection{Discipline-related Differences}
There are disciplines of research which are in general easier for the models to detect,
namely "Physical Sciences" and "History and Archaeology"
which are also the disciplines of research with shorter payloads (word-wise).
An approach to derive a general rule from these differences
is to look for correlations with other variables of the data set:
\begin{itemize}
    \item \textbf{number of words per payload:}
        All $f_{\beta}$-scores correlate negatively with the median number of words per payload (Pearson correlation):
        -0.51 ($f_{0.5}$)
        -0.474 ($f_1$)
        -0.446 ($f_2$).
        These numbers indicate that the usage of a concise vocabulary in the metadata
        improves the chance to be classified correctly.
        In the BoW-approach, disciplines of research with a smaller vocabulary
        have a better relative representation in the set of terms finally selected.
        With regard to the LSTM/embedding approach,
        a possible explanation for the correlation is that shorter text
        are easier to "digest" for the model,
        meaning that further textual content does not improve the performance if it does not contain
        an equivalent semantic surplus.
    \item \textbf{number of labels per payload:}
        The Pearson correlation between the mean number of labels and the $f_{\beta}$-scores is rather weak, and varies in direction:
        -0.093 ($f_{0.5}$)
        0.068 ($f_1$)
        0.083 ($f_2$).
        The number of labels is therefore no explanation for the performance in general, at least not from the aggregated perspective,
        though it is a possible explanation for the outliers:\footnote{The
        same applies to the ratio of 1-labeled payloads to total payloads average number of labels (Pearson correlation):
        0.117 ($f_{0.5}$)
        -0.01 ($f_1$)
        -0.023 ($f_2$).
        }
        In those two cases the amount of shared payloads with other labels is small \emph{and}
        coincides with a semantic distinction.
\end{itemize}

\subsection{Discussion of Miss-Classifications}

This section discusses explanations of the reported miss-classifications and
presents strategies to mitigate the identified problems
or improve the performance by trying out other approaches than presented in this paper.

The following paragraphs focus on different aspects of the presented approach:
\begin{itemize}
    \item data processing.
    \item vectorizing approaches.
    \item model and/or hyper-parameter selection.
\end{itemize}

Each of the following paragraphs focus on one aspect.
While we discuss some possible limitations of our approach for the first two aspects,
the third part is different.
As already stated, the solution space (models + hyper-parameter combinations) is too
vast to be exhaustively searched.
We therefore content ourselves with hints, on how to methodically supersede our results.
This last activity is similar to reproducing our results.

\subsubsection{Data Issues}

There are two promising explanations for miss-classifications
based on a critical review of the training/evaluation set:
\begin{itemize}
    \item Some payloads may be labeled wrong, which means that in such a case the model
        performs better than the person who originally labeled the research data item.
        There are structural explanations available for this assumption that go beyond simple classification errors:
        Some repositories might only allow or encourage one discipline per data set
        or data sets are not curated over time (e.g.\ by adding disciplines after submission).
        Unfortunately, there is no approach known to us to automate a procedure to identify and correct
        such types of miss-classifications other than manual checks:
        \begin{itemize}
            \item Ordering of all false positives by the probability the model assigns to
                the miss-classified labels in decreasing order:
                The higher the probability,
                the more likely is the model's classification correct.
            \item Manual relabeling of the data item if the machine's classification seems warranted.
        \end{itemize}
    \item Some payloads could be of insufficient quality for both model \emph{and} human expert to
        unambiguously classify the research data item.
        This means that the model would be "justified" in a false negative,
        since the payload in question would be too short and/or too unspecific.
        As with the previous idea, there is only a manual approach known to us, to detect and handle such
        cases:
        \begin{itemize}
            \item Ordering of all false negatives by the probability the model assigns to
                the miss-classification and number of words (both in increasing order):
                The lower the probability and shorter the payload,
                the more likely is the model's classification warranted.
            \item Manual assessment of the payload (does it contain enough information
                to make a sound classification decision?)
                If the payload is not found to be sufficient for a classification,
                it can be ignored by future training runs.
        \end{itemize}
\end{itemize}
Both these approaches focus on the quality of the training/evaluation data,
whereas another idea is to enlarge these data by using additional sources,
such as the Bielefeld Academic Search Engine (BASE)\footnote{\url{https://base-search.net}}
or Crossref\footnote{\url{https://crossref.org}}

All three approaches lead to a newer version of the training and evaluation data set,
which necessitates a re-training of the models to assess the improvements by
the baseline presented in this paper.

\subsubsection{Vectorization Issues}
\begin{itemize}
    \item \textbf{BoW}: The large number of input features for the best performing model (100,000)
        slows down prediction and increases the size of the model.
        Aside from these rather technical issues, 3,996,093
        features are not exploited to increase the performance of the models.
        With dimension-reduction mechanisms such as Principal Component Analysis (PCA),
        these unused information might be exploited to further reduce the miss-classifications,
        while the dimension of the input layer of the network is reduced at the same time
        (with beneficial consequences for the prediction latency and model size).
        Another approach is to test
        what impact the selection of another list of stop words would have.
    \item \textbf{Embeddings}: Using another embedding data set to vectorize the payloads
        might allow reduce the number of miss-classifications,
        if these embeddings were trained on a context closer to scientific communication
        than the used Google News data set.
        Another approach is to train the embeddings from scratch or
        update an existing embedding model during the classification process.
\end{itemize}

\subsubsection{Notes on Reproducibility}\label{subsec:reproducibility}

The presented procedure to retrieve, clean, and vectorize the data,
and evaluate different machine learning models on the results
is based on several assumptions, which were motivated in this paper.
In general, the solution space is so vast,
that it is currently unreasonable to check every hyper-parameter combination
or promising learning algorithm considering the resources necessary.

Besides the explanation for the choices we gave in this paper,
we followed an approach that allows to retrace each step taken,
so different configurations and hyper-parameters can be tested.
Nevertheless, to guarantee comparability, it is crucial to keep those
parts of the data processing and the learning pipeline fixed,
which do \emph{not} diverge from our approach.
To ease such a procedure,
the data, source code and all configurations, are made publicly available:
\begin{itemize}
    \item Raw retrieved data: \cite{raw_data}
    \item Cleaned and vectorized training and evaluation data
        \begin{itemize}
            \item small: \cite{small}
            \item medium: \cite{medium}
            \item large: \cite{large}
        \end{itemize}
    \item Source code with sub-modules for the steps presented in this paper,
        \cite{code_config}, with the following components:
        \begin{itemize}
            \item code/retrieve, corresponding to \autoref{subsec:retrieve}
            \item code/clean, corresponding to \autoref{subsec:clean}
            \item code/vectorize, corresponding to \autoref{subsec:vectorize}
            \item code/evaluate, corresponding to \autoref{sec:methods}
            \item config, all configurations, including the stop word list
        \end{itemize}
    \item Statistical data for this paper and evaluation data of all training evaluation runs:
        \cite{paper_data}
\end{itemize}

\subsection{Threats to Validity}\label{subsec:threats}
The decision for a classification scheme is necessarily a political statement.
How borders between disciplines are drawn,
which research activities are aggregated under a label,
and which disciplines are considered to be "neighbors",
should be open to reflection and debate.
The selection of the common classification scheme for this paper,
was steered by technical reasoning to minimize the effort
while maximizing the classifiers' performance (see \autoref{sec:dataHandling}).
The methodological approach and therefore the code,
allows to use another classification scheme,
if mappings are provided from the found schemes to this alternative target scheme.

Another potentially arbitrary feature of the presented approach lies both
in the subject classifications found in the raw data and
in the process to streamline them during the cleaning step.
DataCite aggregates over many sources and therefore over many curators and scientist who make the first type of decision;
\cite{issi005} reports that 762 world-wide organizations were included
as data centers in the DataCite index in April 2016.
This bandwidth hopefully leads to a situation where prejudice and error is averaged out.
Even if some cultural and socialized patterns in classification remain
and some classification schemes are more prominent than others (ANZSRC, DDC),
this approach is to our knowledge the best available.
By filtering out metadata which were clearly classified by other automatic means
(e.g.\ the linsearch classification scheme),
we hope to minimize an amplifier effect
(models trained on data classified by other models).
We managed the second type of arbitrariness
(mapping the found labels to a common scheme),
by making each of the 609,524 mapping decisions
transparent and reproducible,
so that possibly existing biases and uncatched mistakes are corrigible.

\section{Conclusion \& Outlook}\label{sec:conclusion}

In this paper a report is given how training and evaluation metadata describing research data
are retrieved, cleaned, labeled and vectorized,
in order to test the best machine learning model to classify the metadata
by the discipline of research of the research data.
Since this is a multi-label problem,
both training and evaluation procedure must be aligned with technical best practices,
such as stratified sampling or using macro-averaged scores.

MLP models and LSTM models perform well enough for usage in the context of scientometric
research,
while the usage of the evaluated models in assistant systems and value-adding services of
research data providers should only be considered after user interaction studies.
Ideas for improvement of the performance are mostly targeted
at the training and evaluation data.

Ideas for scientometric application of the models on data sources such as
DataCite and BASE include but are not limited to:
\begin{itemize}
    \item Is research becoming more or less interdisciplinary?
        This question can be answered by using the automated classifiers on large time-indexed data sets
        (such as DataCite, which includes the year of publication as a mandatory field).
        The classified data sets will display a trend, whether the number of labels increase,
        decrease or stay stable.
    \item How does each discipline contribute to the growth of research data?
        This question can be answered by analyzing data sources such as DataCite or BASE,
        after their contents have been classified by the presented models.
\end{itemize}
This paper provides the mean to quantify the expected error in such investigations,
based on the reported $f_{\beta}$-scores.

It is furthermore an open question how models honoring the hierarchical nature of
most classification schemes could be trained and implemented at scale
for the use cases at hand.

\section*{Authors' Contribution}
This section follows the Contributor Roles Taxonomy (CRediT).

Tobias Weber:
\begin{itemize}
    \item Conceptualization
    \item Data curation
    \item Formal analysis
    \item Investigation
    \item Methodology
    \item Software
    \item Visualization
    \item Writing - original draft
    \item Writing - review \& editing
\end{itemize}

Dieter Kranzlmüller:
\begin{itemize}
    \item Funding acquisition
    \item Resources
    \item Supervision
\end{itemize}

Michael Fromm:
\begin{itemize}
    \item Methodology
    \item Validation
    \item Writing - review \& editing
\end{itemize}

Nelson Tavares de Sousa:
\begin{itemize}
    \item Conceptualization
    \item Writing - review \& editing
\end{itemize}

\section*{Acknowledgments}
This work was supported by the DFG (German Research Foundation) with the GeRDI project (Grants No. BO818/16-1 and HA2038/6-1). This work has also been funded by
the DFG within the project Relational Machine Learning for Argument Validation (ReMLAV), Grant Number SE 1039/10-1,
as part of the Priority Program "Robust Argumentation Machines (RATIO)" (SPP-1999).

We, as the authors of this work, take full responsibilities for its content;
nevertheless, we want to thank the LRZ Compute Cloud team for providing the resources to run our
experiments,
Martin Fenner from the DataCite-team for comments on an early draft of this paper,
the staff at the library of the Ludwig-Maximilians-Universität München (esp. Martin Spenger),
for support in questions related to the library sciences,
and the Munich Network Management Team for comments and suggestions on the ideas of this paper.

\section*{Competing Interest}
The authors have no competing interests to report.

\printbibliography

\end{document}